\documentclass[twocolumn,showpacs,amsmath,amssymb]{revtex4}
\usepackage{graphicx}
\usepackage{color}

\begin{document}

\title{On Identification of Heat and Work in Quantum Many-Body Systems with Local Operations and Classical Communication}

\author{Hiroaki Matsueda\thanks{hiroaki.matsueda.c8@tohoku.ac.jp}${}^{1,2}$}
\affiliation{${}^{1}$Department of Applied Physics, Graduate School of Engineering, Tohoku University, Sendai 980-8579, Japan}
\affiliation{${}^{2}$Center for Science and Innovation in Spintronics, Tohoku University, Sendai 980-8577, Japan}

\date{\today}
\begin{abstract}
How we identify heat and work is a fundamental question in modern quantum thermodynamics. Usually, heat and work are attributed to changes in the density matrix and the Hamiltonian, respectively, during time-evolution processes in quantum systems. Recently, it has been recognized that this identification is ambiguous. Furthermore, quantum thermodynamics involving quantum measurement is still under development. Motivated by these on-going works, we consider a quantum many-body system from which we extract energy by local operations and classical communication (LOCC) according to the quantum energy teleportation (QET) protocol. The central idea to define heat and work unambiguously is based on a sharp insight into the optimization condition of LOCC. When LOCC is optimized, the extractable energy by QET becomes a daemonic ergotropy; thus, it can be attributed as work. On the other hand, when LOCC is not optimized, we have not squeezed out all the energy with the unitary operation. It means there is uncontrollable energy left in the system. The uncontrollable energy can be attributed as heat after careful treatment of many-body interactions. The heat term consists of nonlocal correlation due to communication between remote participants, and the correlation cannot be directly observed for the participant in the subsystem. Thus, this feature is consistent with the traditional perspective of heat as an uncontrollable energy. To deeply understand the nature of heat, we derive two types of generalized Clausius inequality in our effective quantum thermodynamics, and discuss the direction of the inequality. To justify our perspective, we examine a one-dimensional Kitaev-like model and discuss the physical meaning of effective temperature in our thermodynamics.
\end{abstract}

\maketitle

\section{Introduction}

The identification of heat and work is one of the long-standing problems in quantum thermodynamics~\cite{Strasberg,Potts}. According to the first law of traditional thermodynamics, a change in the energy of a system $dE$ is given by the sum of the heat absorbed by the system $\delta Q$ and the work performed on the system $\delta W$: $dE=\delta Q+\delta W$, where the index $\delta$ represents a path-dependent quantity. In quantum cases, the average internal energy of a quantum system at time $t$ is defined by $\left<E(t)\right>={\rm tr}\left(\rho(t)H(t)\right)$, and the time derivative of this equation naturally contains two different terms: $d\left<E_{A}(t)\right>={\rm tr}\left(d\rho(t)\cdot H(t)\right)+{\rm tr}\left(\rho(t)\cdot dH(t)\right)$. Then, the first (second) term is usually attributed to heat (work). However, this is a very intuitive approach, and recently the limitation of this identification has been widely recognized. In Ref.~\cite{Ahmadi}, the authors found that a change in the density matrix has two contributions and it is better to consider that one of them is attributed to work. The main feature of this new definition is that quantum coherence does not allow all the energy exchanged between two systems to be only of the heat form. Therefore, it is really important to find new ways to define heat that do not depend on any changes in the density matrix. Various studies explain the importance of quantum correlation and coherence in quantum thermodynamics~\cite{Ahmadi,Esposito,Allahverdyan,Hilt,Carrega,Alipour,Xu,Bera,Llobet,Strasberg2,Micadei,Sapienza,Dolatkhah,Clesser,Vallejo,Huang,Holdsworth,Elouard,Dann,Bartosik,Ye,Aguilar,Rivas,Colla,Seegebrecht,Colla2,Rupush,Zhou}. For example, in Ref.~\cite{Esposito}, the entropy production is bounded by the information-geometric distance between the mixed and separable states in a combined system (subsystem-reservoir). This exactly shows that quantum correlation plays a key role in defining thermodynamical quantities.

Quantum-state change by local operations and classical communication (LOCC) is a fundamental procedure in modern quantum thermodynamics. The state change by LOCC is closely related to the change in the energy expectation value for a given Hamiltonian. Thus, LOCC naturally has potential for energy extraction and battery. The unique concepts associated with this storyline are ergotropy and its daemonic generalization~\cite{Allahverdyan2,Pusz,Skrzypczyk,Llobet2,Francica,Bernards,Francica2,Lobejko,Salvia,Touil,Hadipour,Basu}. Although these concepts focus on work extraction from a pure quantum system without environmental degrees of freedom, LOCC is sometimes performed on a subsystem, leading to emergent thermodynamical properties after tracing over the degree of freedom of the rest of the whole system. To the best of my knowledge, the proper definition of heat in that case is still an open question. Furthermore, LOCC is instantaneous and is far from adiabatic conditions. Thus, it is an important topic in non-equilibrium thermodynamics. In addition, many quantum protocols with LOCC consider feedback operation and trajectory averaging. By these processes, the information is necessarily blurred by the averaging procedure, leading to irreversibility. Then, the entropy production or the reverse direction of the Clausius inequality is also an important topic in this research.

These studies are not just of theoretical interest, but fundamental theory that could support future quantum infrastructure, uncovering how to transfer energy between distant quantum computers or quantum batteries without generating heat or with minimal loss. As a counterpart to extracting energy efficiently, dealing with the problem of waste heat is really meaningful.

Towards a complete description of quantum thermodynamics, deriving an exact definition of heat is an important issue. For this purpose, we consider a quantum many-body system from which we extract energy by LOCC according to the quantum energy teleportation (QET) protocol~\cite{Hotta,Hotta2,Hotta3,Hotta4,Frey,Trevison,Rodriguez,Ikeda,Wang,Hotta5,Matsueda,Itoh}. This protocol is a good model that includes all the essential concepts for advancing this project. The protocol is defined for a quantum mechanical system at zero temperature, but since the participants, Alice (daemon) and Bob (who extracts energy), are spatially separated, it is possible to introduce effective thermodynamics for Bob's subsystem after tracing over degrees of freedom of the rest of the whole system ($B$ degrees of freedom, environment). In our previous studies, we have derived the second law of information thermodynamics for the energy reduction of Bob's subsystem~\cite{Matsueda,Itoh,Sagawa,Tajima,Park,Funo,Manzano,Minagawa}. These results clearly tell us the critical role of a thermodynamic view of the QET protocol, even though it is formulated at zero temperature. In the present work, we focus on the optimization condition of LOCC to properly define heat in our effective thermodynamics. When the LOCC is optimized, the extractable energy by QET becomes a daemonic ergotropy, and thus it can be attributed as work in terms of modern quantum thermodynamics. On the other hand, when the LOCC is not optimized, we have not squeezed out all the energy with the unitary operation. This means that there is uncontrollable energy left in the system. The central idea in this study is to regard this uncontrollable energy as a crucial source of heat. On the basis of this idea, we derive several thermodynamic relations. In particular, we derive two types of Clausius inequality best customized for the QET protocol. After that, we arrive at the proper definition of heat. The heat term consists of nonlocal correlations between Alice and Bob, and the correlation cannot be directly observed by Bob. Thus, this feature is consistent with the traditional perspective of heat as an uncontrollable energy. In addition to the proper definition of heat, it is quite interesting to explore the direction of the Clausius inequality in our effective quantum thermodynamics. We will discuss the nature of the inequality and the physical meaning of effective inverse temperature by using a one-dimensional Kitaev-like model.

The organization of this paper is as follows. In Sec.~\ref{QET}, we introduce the energy extraction protocol by QET. In Sec.~\ref{Duality}, we derive the proper definition of work and heat from a sharp insight into the daemonic ergotropy. In Sect.~\ref{DerivationofCI}, we derive a generalized Clausius inequality and examine its typical properties. In Sec.~\ref{model}, we introduce a Kitaev-like one-dimensional model that has already been examined in the previous paper~\cite{Matsueda}, and apply our thermodynamic perspectives to this model. We check the tightness of the Clausius inequality, and particularly discuss the meaning of effective inverse temperature. Finally, we summarize our study in Sec.~\ref{summary}.

\section{Energy Extraction Protocol by LOCC (QET)}
\label{QET}

\begin{figure}[htbp]
\begin{center}
\includegraphics[width=7cm]{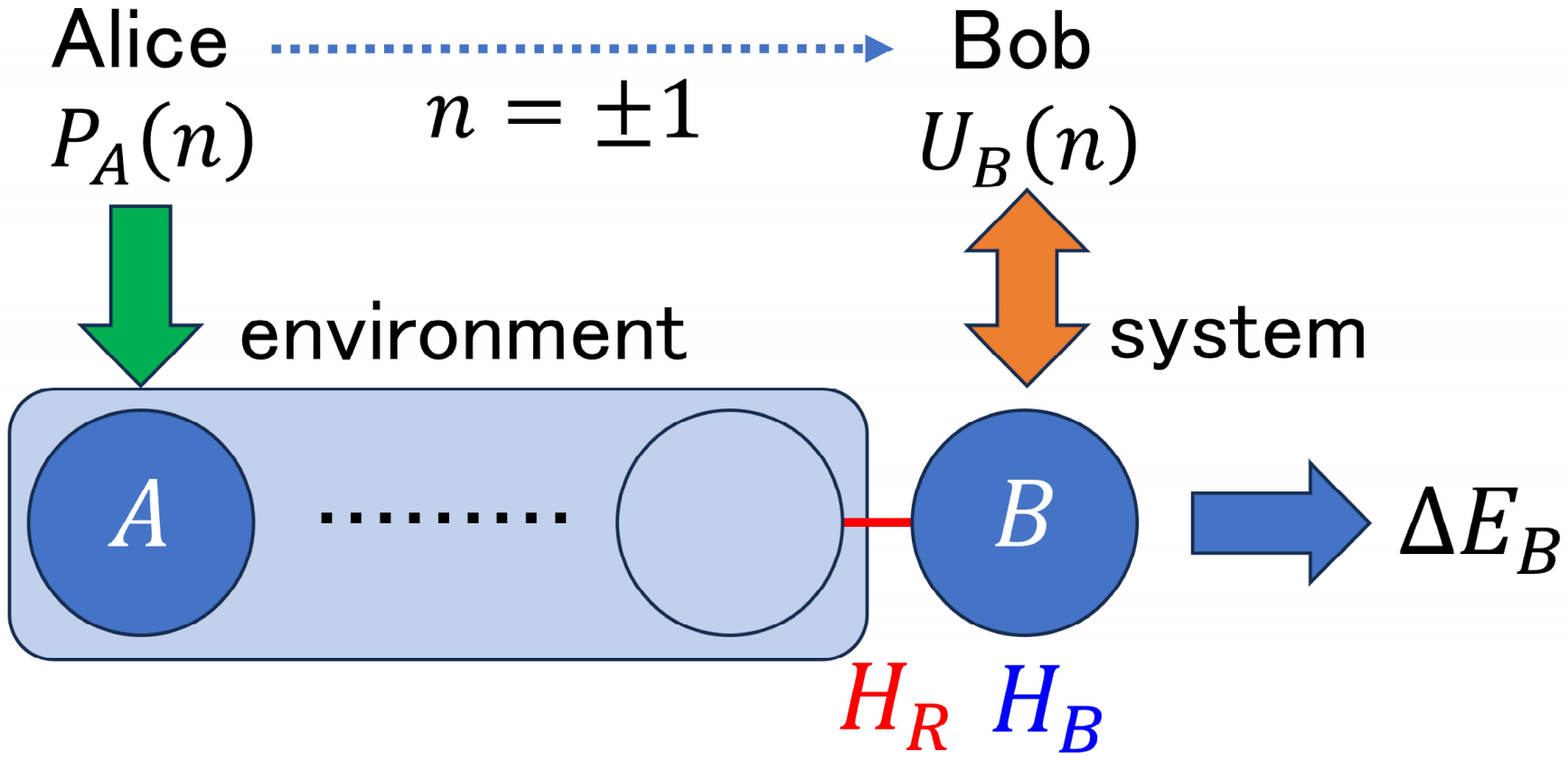}
\end{center}
\caption{Thermodynamic view of our model and protocol. The total system is a energy-extraction device, and Bob operates local feedback unitary to a local part of the total system depending on the Alice's measurement result. Then, the energy reduction may occur near the Bob's local system. The rest of the total system acts as an environment for Bob, and Bob cannot directly access to the environment.
}
\label{WorkHeatfig1}
\end{figure}

Consider remote local quantum systems (two single spins, for example) $A$ and $B$ coupled through the interaction term $V$. The interaction term contains many sites. The Hamiltonian is given by $H=H_{A}+V+H_{B}$. Our aim is to extract positive energy from a local system $B$ through LOCC. For this purpose, Alice performs a projective measurement $P_{A}(n)$ on a local system $A$, and subsequently the measurement result (the result is sometimes called an outcome or a trajectory) $n$ ($=\pm 1$) is sent to Bob. Bob performs a feedback unitary transformation $U_{B}(n)$, to the local system $B$ to extract energy outside of the whole system. Figure~\ref{WorkHeatfig1} shows thermodynamical view of our model and protocol, where because of Bob's local operation the rest of the whole system behaves as an environment.

We evaluate the energy injected into the system by Alice's measurements, $\Delta E_{A}=E_{A}-\epsilon$ ($\epsilon$ is the ground-state energy of the given Hamiltonian $H$ and $H\left|\psi\right>=\epsilon\left|\psi\right>$), and the energy extracted by Bob, $\Delta E_{B}=E_{A}-E_{B}$. The energy expectation value after Alice's measurement is defined by
\begin{align}
E_{A}=\sum_{n}p_{n}\left<\psi_{A}(n)\right| H\left|\psi_{A}(n)\right> ,
\end{align}
where the state after Alice's measurement is defined by $\left|\psi_{A}(n)\right>=P_{A}(n)\left|\psi\right>/\sqrt{p_{n}}$ and the probability of finding the measurement result $n$ is given by $p_{n}=\left<\psi\right|P_{n}(n)\left|\psi\right>$. Here, the expectation value is averaged over all possible trajectories. The energy injected by Alice, $\Delta E_{A}$, is given by
\begin{align}
\Delta E_{A}=\sum_{n}p_{n}\left<\psi_{A}(n)\right|H_{A}+V\left|\psi_{A}(n)\right>-\left(\epsilon_{A}+\epsilon_{V}\right) ,
\end{align}
where $\epsilon_{A}=\left<\psi\right|H_{A}\left|\psi\right>$ and $\epsilon_{V}=\left<\psi\right|V\left|\psi\right>$. We also introduce the energy expectation value after Bob's feedback unitary operation:
\begin{align}
E_{B}=\sum_{n}p_{n}\left<\psi_{A}(n)\right|U_{B}^{\dagger}(n)HU_{B}(n)\left|\psi_{A}(t)\right> .
\end{align}
Due to the fact that $U_{B}(n)$ acts on the system quite locally and commutes with $H_{A}$, $E_{B}$ can be transformed into the following form:
\begin{align}
E_{B}=&\sum_{n}p_{n}\left<\psi_{A}(n)\right|H_{A}\left|\psi_{A}(n)\right> \nonumber \\
&+\sum_{n}p_{n}\left<\psi_{A}(n)\right|U_{B}^{\dagger}(n)\left(V+H_{B}\right)U_{B}(n)\left|\psi_{A}(n)\right> .
\end{align}
The energy extracted by Bob, $\Delta E_{B}$, is finally given by
\begin{align}
\Delta E_{B}=\Delta E_{B,B}+\Delta E_{B,R} , \label{first}
\end{align}
where $\Delta E_{B,i}$ ($i=B,R$) are defined respectively by
\begin{align}
\Delta E_{B,i}&=\epsilon_{i}-\sum_{n}p_{n}\left<\psi_{A}(n)\right|U_{B}^{\dagger}(n)H_{i}U_{B}(n)\left|\psi_{A}(n)\right> \nonumber \\
&=\epsilon_{i}-\sum_{n}\left<\psi\right|P_{A}(n)U_{B}^{\dagger}(n)H_{i}U_{B}(n)\left|\psi\right>, \label{DEBi}
\end{align}
with $\epsilon_{B}=\left<\psi\right|H_{B}\left|\psi\right>$ and $\epsilon_{R}=\left<\psi\right|H_{R}\left|\psi\right>$. Here, $H_{R}$ (interaction between spin $B$ and its neighboring spin) is a rightmost part of $V$, and does not commute with $U_{B}(n)$. If $\Delta E_{B}$ becomes positive, Bob can extract energy outside the system.

\section{Daemonic Ergotropy and Optimization of LOCC: Towards Proper Definition of Heat}
\label{Duality}

The QET protocol itself does not specify the quality of energy such as work and heat. The precise definition of practical work should be described by ergotropy or daemonic ergotropy in modern quantum thermodynamics~\cite{Allahverdyan2,Pusz,Skrzypczyk,Llobet2,Francica,Bernards,Francica2,Lobejko,Salvia,Touil,Hadipour,Basu}. Ergotropy is defined by the energy difference between target state $\rho$ and its passive state $U\rho U^{\dagger}$: ${\mathcal E}\left(\rho\right)={\rm tr}\left(\rho H\right)-\min_{U}{\rm tr}\left(U\rho U^{\dagger}H\right)$. The daemonic ergotropy is an extension of the ergotropy under LOCC with the concept of Maxwell's daemon. For the combined state $\rho_{AB}$ ($A$ and $B$ denote the ancilla and system, respectively), a daemon measures the ancilla by a set of operators $\left\{P_{n}\right\}$, and then the system shrinks into the state $\rho_{B|n}$ according to the measurement result. Then, the daemonic ergotropy is defined by ${\mathcal E}_{D}\left(\rho_{AB}\right)=\max_{\left\{P_{n}\right\}}\sum_{n}p_{n}{\mathcal E}\left(\rho_{B|n}\right)$, where $p_{n}$ is the measurement probability to find the result $n$. It has been proved that ${\mathcal E}_{D}\left(\rho_{AB}\right)-{\mathcal E}\left(\rho_{B}\right)\ge 0$ and the difference is called daemonic gain~\cite{Francica,Bernards}.

The extractable energy by QET is represented as
\begin{align}
\Delta E_{B}=\sum_{n}p_{n}{\rm tr}\left(\rho^{\rm m}(n)\left(H-U_{B}^{\dagger}(n)HU_{B}(n)\right)\right),
\label{DEBdaemon}
\end{align}
where the density matrix after Alice's projective measurement with result $n$, $\rho^{\rm m}(n)$, is defined by
\begin{align}
\rho^{\rm m}(n)=\left|\psi_{A}(n)\right>\left<\psi_{A}(n)\right|,
\end{align}
and {\it after optimization of $P_{A}(n)$ and $U_{B}(n)$} toward the maximum of $\Delta E_{B}$, it seems to be a kind of daemonic ergotropy for spatially distant participants. In this sense, $\Delta E_{B}$ is an extractable work enhanced by daemonic gain. An important view for its generalization is to consider cases in which $\Delta E_{B}$ is {\it not} maximized. Then, Bob does not squeeze out all the energy with his feedback unitary operations, indicating that there is uncontrollable energy left in the whole system. This uncontrollable energy should be assigned as a crucial source of heat, if we can introduce a thermodynamical view (at the present stage, we just consider quantum mechanics at zero temperature). Thus, it is reasonable to introduce the following
\begin{align}
q\left(P_{A}(n),U_{B}(n)\right)=\Delta E_{B}\left(P_{A}(n),U_{B}(n)\right)-\Delta E_{B}^{\rm max},
\label{defq}
\end{align}
as the source of heat, and the maximum condition of $\Delta E_{B}$, $\Delta E_{B}=\Delta E_{B}^{\rm max}$, is simply represented as $q\left(P_{A}^{\ast}(n),U_{B}^{\ast}(n)\right)=0$, where $P_{A}^{\ast}(n)$ and $U_{B}^{\ast}(n)$ are Alice's projective measurement and Bob's feedback unitary operators for $\Delta E_{B}^{\rm max}$. We must take special care about the many-body interaction between Bob's subsystem and the rest of the whole system. After that, we will derive the rigorous definition of heat $Q$. In the next section, we will discuss about it.

Hereafter, we will use several notations for the maximum value of $\Delta E_{B}$ and the corresponding LOCC, $\Delta E_{B}^{\rm max}=\Delta E_{B}^{\ast}=\Delta E_{B}\left(P_{A}^{\ast}(n),U_{B}^{\ast}(n)\right)=\Delta E_{B}\left(\vec{r}^{\ast},\vec{s}^{\ast},\theta^{\ast}\right)$, depending on the calculation details. Here, $q$ is defined so that it is always non-positive, $q\le 0$. This is because the extractable energy outside the system is defined as a positive quantity and the heat remains in the system.

To examine the nature of the uncontrollable energy, it is helpful to pay attention to Bob's effective thermodynamics. This is because Bob locally performs the feedback unitary operation for his subsystem. The attention is also closely related to our previous observation that the second law of information thermodynamics applies to $\Delta E_{B,B}$ rather than to $\Delta E_{B}$~\cite{Matsueda}. Carefully speaking, there is no net energetic flow across the boundary between Bob's subsystem and the rest of the whole system, since $q$ is almost locally defined from $H_{B}$ and $H_{R}$ and it means that the uncontrollable energy is localized around Bob's subsystem. However, there exists a change in bulk long-range correlation across the boundary during the QET process that can be transformed into virtual heat for an effective temperature, since long-range communication between Alice and Bob and entanglement are essential for QET. Such information that Bob cannot directly access is essential to define heat in Bob's effective thermodynamics. In the next section, we will see that the proper definition of heat $Q$ naturally contains the information.

\section{Generalized Clausius Inequality}
\label{DerivationofCI}

\subsection{Reference Thermal State and the Second Law of Effective Information Thermodynamics}

We examine the nature of $q$ by deriving the corresponding Clausius inequality. This examination provides us with the proper definition of heat $Q$ on the basis of $q$ and the effective inverse temperature $\beta_{\rm eff}$. A method to introduce a temperature-like concept by applying Clausius' equality to Shannon entropy, thereby converting pure quantum mechanics into a thermodynamic description, has been studied before~\cite{Abe1,Abe2,Abe3,Bender}. The Clausius inequality is a fundamental principle in thermodynamics that states that the change in entropy of a system is greater than or equal to the heat transferred to the system divided by the temperature, indicating that entropy tends to increase in irreversible processes. In the QET case, the trajectory average over the measurement results is an information erasure, and then the process effectively becomes irreversible. This feature induces an increase in entropy, relevant to the Clausius inequality. At the same time, the QET protocol proceeds with exhausting the given quantum resource in the ground state. In such a case, the direction of the inequality can be reversed~\cite{Aguilar}. Thus, we will particularly focus on the direction of the Clausius inequality.

For this purpose, we introduce a reference thermal equilibrium state
\begin{align}
\sigma_{B}=\frac{1}{z_{B}}e^{-\beta_{\rm eff}H_{B}}, \label{sigmaB}
\end{align}
and we could say that in the QET protocol, Clausius inequality thermodynamically formulates how far the quantum state being manipulated by LOCC is from $\sigma_{B}$. The inverse temperature $\beta_{\rm eff}$ can be determined so that our Clausius inequality gives a tight and meaningful bound.

In our previous works~\cite{Matsueda,Itoh}, we have derived the second law of information thermodynamics for the upper bound of $\Delta E_{B,B}$, which is represented as
\begin{align}
\Delta E_{B,B}\le\Delta E_{B,B}^{\rm max}=\frac{1}{\beta_{\rm eff}}\left[D\left(\rho_{B}^{\rm i}||\sigma_{B}\right)+I_{\rm QC}\right],
\label{DEBBmax}
\end{align}
where $\rho_{B}^{\rm i}={\rm tr}_{\bar{B}}\left|\psi\right>\left<\psi\right|$, the Kullback-Leiber (KL) divergence (or quantum relative entropy) is defined by $D\left(\rho||\sigma\right)={\rm tr}_{B}\left(\rho\log\rho-\rho\log\sigma\right)$, and the QC-mutual information is defined by $I_{\rm QC}=S\left(\rho_{B}^{\rm i}\right)-\sum_{n}p_{n}S\left(\rho_{B}^{\rm m}(n)\right)$ with $S(\rho)=-{\rm tr}_{B}\left(\rho\log\rho\right)$. The inverse temperature is then determined from the equality condition of Eq.~(\ref{DEBBmax}), and this condition leads to
\begin{align}
S\left(\sigma_{B}\right)=\min_{\left\{P_{A}(n)\right\}}\sum_{n}p_{n}S\left(\rho_{B}^{\rm m}(n)\right). \label{msigma}
\end{align}
If $p_{n}$ and the eigenvalues of $\rho_{B}^{\rm m}(n)$ are independent of the measurement result $n$, we can relate $\rho_{B}^{\rm m}(n)$ with $\sigma_{B}$, and from this relation we can derive $\beta_{\rm eff}$ uniquely.

To derive the second law of effective information thermodynamics, we focus on the consumption of quantum resources such as non-equilibriumness and the entanglement of the ground state by Alice's projective measurement. To derive the Clausius inequality, we focus on the entropy change due to Bob's feedback unitary operation.

\subsection{Derivation of Generalized Clausius Inequality (Type-I and Type-II formulae), Proper Definition of Heat, and Information-Geometrical Interpretation of Inequality}

We derive two useful forms of Clausius inequality, called type-I and type-II formulae in this paper, and provide their quantum information-geometrical interpretation.

A derivation of our Clausius inequality starts from substituting $H_{B}=-\left(\log\sigma_{B}+\log z_{B}\right)/\beta_{\rm eff}$ into Eq.~(\ref{defq}), leading to the following result:
\begin{align}
&\beta_{\rm eff}\left(q-\Delta E_{B,R}+\Delta E_{B,R}^{\ast}\right) \nonumber \\
&\;\; ={\rm tr}_{B}\left(\left(\bar{\rho}_{B}^{{\rm m}\ast}-\bar{\rho}_{B}^{{\rm f}\ast}\right)\log\sigma_{B}\right) \nonumber \\
&\;\;\;\;\;\;-{\rm tr}_{B}\left(\left(\bar{\rho}_{B}^{\rm m}-\bar{\rho}_{B}^{\rm f}\right)\log\sigma_{B}\right),
\label{intermediate}
\end{align}
where we have introduced the following abbreviations:
\begin{align}
\bar{\rho}_{B}^{\rm m}=&\sum_{n}p_{n}\rho_{B}^{\rm m}(n) , \\
\bar{\rho}_{B}^{\rm f}=&\sum_{n}p_{n}U_{B}(n)\rho_{B}^{\rm m}(n)U_{B}^{\dagger}(n),
\end{align}
and we have added the symbol $\ast$ to the quantities for the optimal LOCC process. We next transform the right-hand side (RHS) of Eq.~(\ref{intermediate}) by using the entropy function and the KL divergence. Then, we arrive at the form of Clausius inequality (here we call it type-I formula) given by
\begin{align}
\beta_{\rm eff}Q+\Delta S_{B}=\Delta D, \label{Clausius1}
\end{align}
where $Q$, $\Delta S_{B}$, and $\Delta D$ are defined respectively by
\begin{align}
Q=&q-\Delta E_{B,R}+\Delta E_{B,R}^{\ast}=\Delta E_{B,B}-\Delta E_{B,B}^{\ast} , \label{Q} \\
\Delta S_{B}=&\left[S\left(\bar{\rho}_{B}^{\rm f}\right)-S\left(\bar{\rho}_{B}^{\rm m}\right)\right]-\left[S\left(\bar{\rho}_{B}^{{\rm f}\ast}\right)-S\left(\bar{\rho}_{B}^{{\rm m}\ast}\right)\right] , \label{DSB} \\
\Delta D=&\left[D\left(\bar{\rho}_{B}^{\rm m}||\sigma_{B}\right)-D\left(\bar{\rho}_{B}^{\rm f}||\sigma_{B}\right)\right] \nonumber \\
&-\left[D\left(\bar{\rho}_{B}^{{\rm m}\ast}||\sigma_{B}\right)-D\left(\bar{\rho}_{B}^{{\rm f}\ast}||\sigma_{B}\right)\right]. \label{DD}
\end{align}
We regard Eq.~(\ref{Q}) as the proper definition of heat $Q$. $Q$ shows how far $\Delta E_{B,B}$ at the current LOCC is from $\Delta E_{B,B}^{\ast}$ during the optimal LOCC. We find $Q^{\ast}=0$. Here, we are able to derive $Q$ using the Clausius relation without introducing any changes in the density matrix. Mathematically, it is possible to decompose $H_{R}$ into a product of $H_{B}$ and a composite operator $R_{\bar{B}B}$, and renormalize $R_{\bar{B}B}$ into the change in the density matrix. However, in this case, we always suffer from the 
violation of the spectral positivity of density matrices. For this reason, we believe that $Q$ is uniquely defined by Eq.~(\ref{Q}) even in the cases with many-body interactions between Bob's subsystem and the rest of the whole system. This is a very important perspective. We also find that $\Delta D$ can take both positive and negative values. The type-I formula is conveniently used to examine the violation of the standard Clausius inequality. Equation~(\ref{Clausius1}) is reduced to the standard form of Clausius theorem $-Q=\Delta S_{B}/\beta_{\rm eff}$ for a reversible process, if the quantum-mechanical state change can be omitted. This reduction is reasonable, since the traditional equilibrium thermodynamics does not treat quantum-state change.

There are four important remarks associated with Eqs.~(\ref{Q}), (\ref{DSB}), and (\ref{DD}). The first remark is about the range of $Q$. Clearly, the upper and the lower bounds of $Q$ are represented by the following inequality
\begin{align}
\Delta E_{B,B}^{\rm min}-\Delta E_{B,B}^{\ast}\le Q\le \Delta E_{B,B}^{\rm max}-\Delta E_{B,B}^{\ast}.
\end{align}
Unlike $q$, which only takes negative values, $Q$ can take both positive and negative values ($Q^{\rm max}=\Delta E_{B,B}^{\rm max}-\Delta E_{B,B}^{\ast}\ge 0$, $Q^{\rm min}=(-Q)^{\rm max}=\Delta E_{B,B}^{\rm min}-\Delta E_{B,B}^{\ast}\le 0$). Since $q$ is an uncontrollable energy left in the system, the positive value of $Q$ suggests a heat flow from Bob's subsystem to the reset of the whole system. We have already discussed that there is no bulk flow of energy. Thus, this flow is related to the nonlocal correlation between Alice and Bob. Equation~(\ref{Q}) contains both heat $Q$ and work $\Delta E_{B,B}$, but it may not be correct that we consider Eq.~(\ref{Q}) as the first law of our effective thermodynamics. This is because it is not clear whether there exist locally conserved quantities defined in Bob's subsystem.

The next remark is about roles of Alice's projective measurement in heat generation~\cite{Elouard2,Yamamoto}. According to Ref.~\cite{Yamamoto}, where the authers treated a qubit coupled to heat bathes under continuous quantum measurement, they found that heat always flows from the measurement apparatus into the qubit. In the definition of $Q$ in Eq.~(\ref{Q}), Alice's projective measurement operator $P_{A}(n)$ takes different values in $\Delta E_{B,B}$ and $\Delta E_{B,B}^{\ast}$, and the projective measurement may generate heat. Since $Q$ is represented as $\Delta E_{B,B}$, we should think of $U_{B}(n)$ as the priority source of heat generation. To resolve this question, we may need to completely separate the roles of $P_{A}(n)$ and $U_{B}(n)$.

The third remark is about the range of $\Delta S_{B}$. We can prove the following inequality
\begin{align}
&-H\left(p_{n}\right)\le S\left(\sum_{n}p_{n}U_{B}(n)\rho_{B}^{\rm m}(n)U_{B}^{\dagger}(n)\right) \nonumber \\
&\;\;\;\;\;\;\;\;\;\;\;\;\;\;\;\;\;\;\;-S\left(\sum_{n}p_{n}\rho_{B}^{\rm m}(n)\right)\le H\left(p_{n}\right),
\end{align}
where $H\left(p_{n}\right)=-\sum_{n}p_{n}\log p_{n}$ is a classical Shannon entropy. For this derivation, we use the unitary invariance of von Neumann entropy, concavity $S\left(\sum_{n}p_{n}\rho_{B}^{\rm m}(n)\right)\ge\sum_{n}p_{n}S\left(\rho_{B}^{\rm m}(n)\right)$, and the upper limit of entropy increase by mixture $S\left(\sum_{n}p_{n}\rho_{B}^{\rm m}(n)\right)\le \sum_{n}p_{n}S\left(\rho_{B}^{\rm m}(n)\right)+H\left(p_{n}\right)$. Then, $\Delta S_{B}$ is bounded as
\begin{align}
-H\left(p_{n}\right)-H\left(p_{n}^{\ast}\right)\le\Delta S_{B}\le H\left(p_{n}\right)+H\left(p_{n}^{\ast}\right).
\end{align}
Thus, $\Delta S_{B}$ takes both positive and negative values. This result partially supports the presence of negative $\Delta D$ cases.

\begin{figure}[htbp]
\begin{center}
\includegraphics[width=6cm]{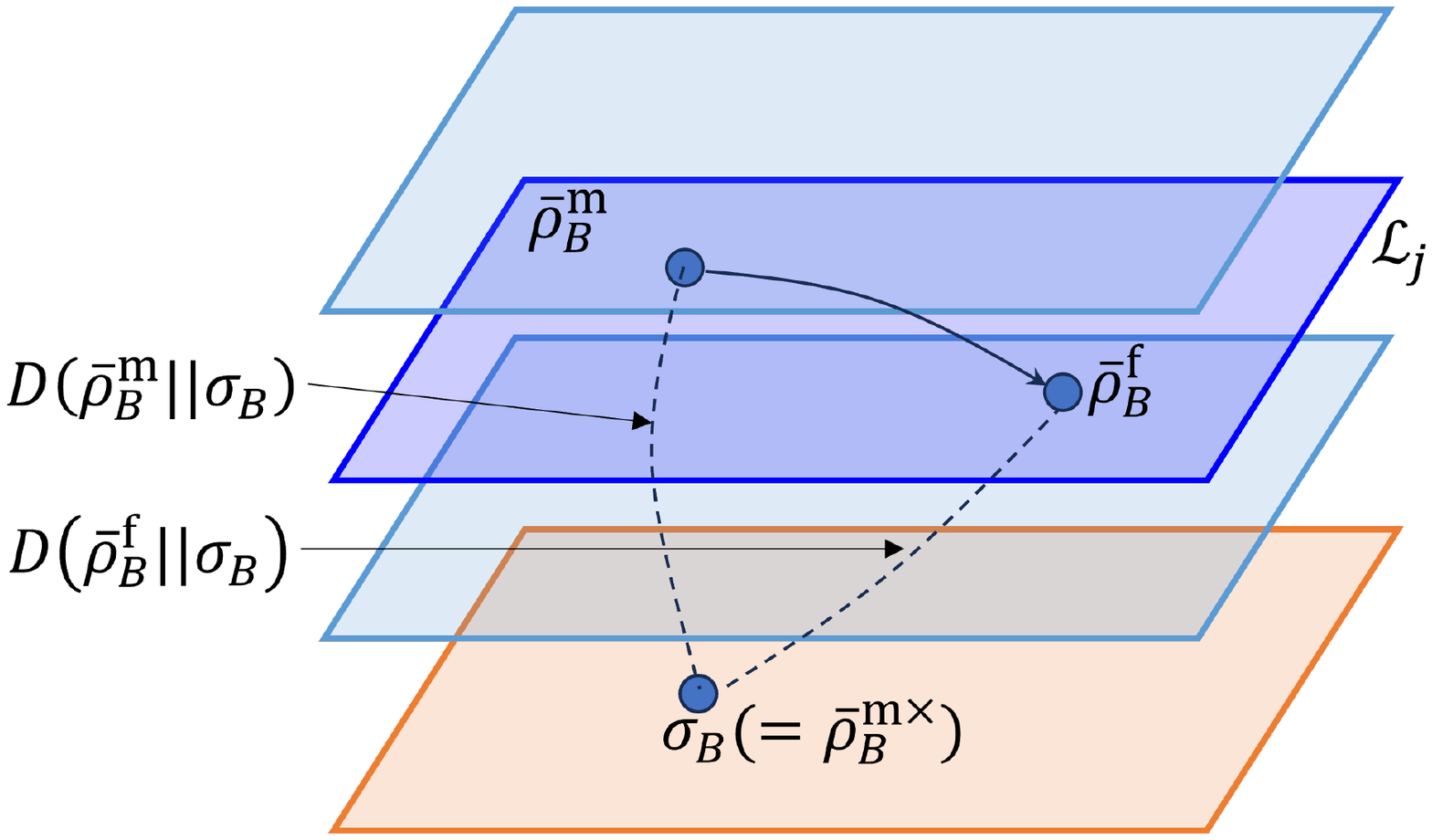}
\end{center}
\caption{Geometrical relation among $\bar{\rho}_{B}^{\rm m}$, $\bar{\rho}_{B}^{\rm f}$, and $\sigma_{B}$ in terms of quantum information geometry. The total manifold $\cal M$ characterizing the quantum state space is foliated one, and is decomposed into a set of leafs $\left\{{\cal L}_{j}\right\}$. The leaf ${\cal L}_{j}$ is labeled by a particular choice of control parameters in $P_{A}(n)$ and $U_{B}(n)$ for successful energy extraction. [Later, we will see that in the Kitaev-like model case we take $j=\left(\vec{r},\vec{s}\right)$]. The reference state $\sigma_{B}$ is taken so that it belongs to the leaf where the heat becomes maximum.
}
\label{WorkHeatfig2}
\end{figure}

The final remark is about an information-geometrical meaning in Eq.~(\ref{DD}) (see Fig.~\ref{WorkHeatfig2}). The difference of two KL divergences in $\Delta D$ is further transformed into
\begin{align}
&D\left(\bar{\rho}_{B}^{\rm f}||\sigma_{B}\right)-D\left(\bar{\rho}_{B}^{\rm m}||\sigma_{B}\right) \nonumber \\
&\;\; =D\left(\bar{\rho}_{B}^{\rm f}||\bar{\rho}_{B}^{\rm m}\right)+{\rm tr}_{B}\left(\left(\bar{\rho}_{B}^{\rm f}-\bar{\rho}_{B}^{\rm m}\right)\left(\log\bar{\rho}_{B}^{\rm m}-\log\sigma_{B}\right)\right).
\label{geometric}
\end{align}
The first term of the RHS in Eq.~(\ref{geometric}) corresponds to the information-geometrical distance between $\bar{\rho}_{B}^{\rm m}$ and $\bar{\rho}_{B}^{\rm f}$. To evaluate the second term of the RHS in Eq.~(\ref{geometric}), let us consider the state space in terms of dual connection in information geometry. If state $\bar{\rho}_{B}^{\rm m}$ is the point projected from state $\sigma_{B}$ along the $e$-geodesic (exponential geodesic) onto some $m$-flat (mixed flat) subspace that includes state $\bar{\rho}_{B}^{\rm f}$, the second term becomes zero. This is a quantum version of Pythagoras theorem. More generally, if a subset of $\cal{M}$, ${\cal L}_{j}$, is $m$-convex and state $\rho$ is the point satisfying $\rho=\arg\min_{\rho^{\prime}\in{\cal L}_{j}}D\left(\rho^{\prime}||\sigma_{B}\right)$, then ${\rm tr}_{B}\left(\left(\bar{\rho}_{B}^{\rm f}-\rho\right)\left(\log\rho-\log\sigma_{B}\right)\right)\ge 0$ for $\forall\bar{\rho}_{B}^{\rm f}\in{\cal L}_{j}$. This quantum state space can be seen as a foliated manifold, and each leaf ${\cal L}_{j}$ is characterized by a set of control parameters $j$ included in $P_{A}(n)$ and $U_{B}(n)$ for successful energy extraction. Bob's feedback unitary operation corresponds to a process along a $m$-geodesic. In the later discussion, $\sigma_{B}$ is set to belong to the leaf with the parameter set when the heat becomes maximum.

The description that treats geodesics in quantum state space as processes of optimal operations can also be seen in Grover's quantum search algorithm and others~\cite{Cafaro}. Moreover, in recent years, there have been methods that study thermodynamic uncertainty using optimal transport theory~\cite{Vu,Nakazato,Dechant}.

Next, let us derive the type-II formula. Although the type-I formula is convenient for detecting the sign of $\Delta D$, it is not necessarily suitable for evaluating the amount of heat itself. The type-II formula is suitable for this purpose. According to our previous works~\cite{Matsueda,Itoh}, it was important to consider the equality between $\sigma_{B}$ and $\bar{\rho}_{B}^{\rm m}$ to determine the effective inverse temperature $\beta_{\rm eff}$. On the basis of this result, we consider the non-negativity of $D\left(\bar{\rho}_{B}^{\rm m}||\sigma_{B}\right)$ in the RHS of Eq.~(\ref{DD}). Then, we arrive at the type-II formula
\begin{align}
-Q\le\Pi, \label{Clausius2}
\end{align}
where the upper bound $\Pi$ is defined by
\begin{align}
\Pi=&\frac{1}{\beta_{\rm eff}}\left[\Delta S_{B}+D\left(\bar{\rho}_{B}^{\rm f}||\sigma_{B}\right)\right. \nonumber \\
&\;\;\;\;\;\;\;\;\left.+\left(D\left(\bar{\rho}_{B}^{{\rm m}\ast}||\sigma_{B}\right)-D\left(\bar{\rho}_{B}^{{\rm f}\ast}||\sigma_{B}\right)\right)\right] \nonumber \\
=&\epsilon_{B}+\frac{1}{\beta_{\rm eff}}\left(\log z_{B}-S\left(\bar{\rho}_{B}^{\rm m}\right)\right) \nonumber \\
&+{\rm tr}_{B}\left(\bar{\rho}_{B}^{\rm f}H_{B}\right)-{\rm tr}_{B}\left(\bar{\rho}_{B}^{{\rm f}\ast}H_{B}\right). \label{Pi}
\end{align}
The present result reminds us of $\Delta E_{B,B}^{\rm max}$ in Eq.~(\ref{DEBBmax}) that contains both $D\left(\rho_{B}^{\rm i}||\sigma_{B}\right)$ and $I_{\rm QC}$.

\subsection{Maximal Heat: Tight Upper Bound of Clausius Inequality}

In our theory, we define thermodynamic quantities based on the deviation from the optimal LOCC. To examine heat further, we to focus especially on cases where a lot of heat is wasted.

In this respect, we consider the equality condition of Eq.~(\ref{Clausius2}). This is given by $\sigma_{B}=\bar{\rho}_{B}^{\rm m}$. Since $\bar{\rho}_{B}^{\rm m}$ is already in thermal equilibrium, we can say that the QET protocol is the process of moving away from it. In this case, the upper bound of Eq.~(\ref{Clausius2}) becomes tight, and we obtain $-Q=\Pi$. The equality condition $\sigma_{B}=\bar{\rho}_{B}^{\rm m}$ is also consistent with the stationary condition of $\Pi$, $\partial\Pi/\partial\beta_{\rm eff}=0$. Thus, $\sigma_{B}=\bar{\rho}_{B}^{\rm m}$ is exactly the minimization condition of $\Pi$, and it also determines the effective inverse temperature, $\sigma_{B}\left(\beta_{\rm eff}\right)=\bar{\rho}_{B}^{{\rm m}\times}$. Here the symbol $\times$ denotes the parameter choice of $\vec{r}=\vec{r}^{\times}$, $\vec{s}=\vec{s}^{\times}$, and $\theta=\theta^{\times}$ so that $-Q$ is maximized. The tightness requires
\begin{align}
(-Q)^{\rm max}=\Pi^{\rm min}\left(\beta_{\rm eff}\right). \label{Qmax}
\end{align}
In this case, $\Pi^{\rm min}\left(\beta_{\rm eff}\right)$, is transformed into
\begin{align}
\Pi^{\rm min}\left(\beta_{\rm eff}\right)
={\rm tr}_{B}\left(\bar{\rho}_{B}^{{\rm f}\times}H_{B}\right)-{\rm tr}_{B}\left(\bar{\rho}_{B}^{{\rm f}\ast}H_{B}\right).
\label{Pimin}
\end{align}
We find that the maximum heat is given by the difference between the local energy of the final state under optimal LOCC and the local energy of the final state under general LOCC. The result seems to be quite reasonable.

The condition $\sigma_{B}=\bar{\rho}_{B}^{\rm m}$ is consistent with Eq.~(\ref{msigma}) and corresponds to the maximization condition of $\Delta E_{B,B}$. The result suggests that the maximization condition of $\Delta E_{B,B}$ is different from that of $\Delta E_{B}$. In our previous paper~\cite{Matsueda}, we have actually found that although $\Delta E_{B,B}$ is maximized, $\Delta E_{B,R}$ takes a large negative value, leading to failure in extracting energy ($\Delta E_{B}<0$). In the viewpoint of heat generation due to Bob's feedback unitary operation, the maximization process of $\Delta E_{B,B}$ generates a lot of heat, and thus the energy extraction is not successful.

Equation~(\ref{Clausius2}) in terms of the type-I form is represented as
\begin{align}
&\beta_{\rm eff}Q+\Delta S_{B} \nonumber \\
&\ge -D\left(\bar{\rho}_{B}^{\rm f}||\sigma_{B}\right)-D\left(\bar{\rho}_{B}^{{\rm m}\ast}||\sigma_{B}\right)+D\left(\bar{\rho}_{B}^{{\rm f}\ast}||\sigma_{B}\right).
\end{align}
When we consider the optimal LOCC, $\bar{\rho}_{B}^{\rm f}=\bar{\rho}_{B}^{{\rm f}\ast}$, the RHS of this equation is reduced to $-D\left(\bar{\rho}_{B}^{{\rm m}\ast}||\sigma_{B}\right)$ and this is clearly negative. In this case, $Q=\Delta S_{B}=0$, and the bound is not tight. However, the negativity suggests that the entanglement is efficiently used for energy extraction~\cite{Aguilar}.

In the next section, we will introduce a QET model to examine heat in more detail. The aforementioned case is useless in terms of high QET performance, but it is also meaningful to understand the worst case to avoid it.

\section{A QET Model}
\label{model}

\subsection{Kitaev-like one-dimensional model and its QET Performance}

We introduce a QET model to understand the nature of $Q$. A nice example is given in our previous paper in which a Kitaev-like one-dimensional model is introduced~\cite{Matsueda,Kitaev,Nussinov,Hermanns}. The Hamiltonian is given by $H=H_{A}+V+H_{B}=H_{A}+H_{L}+H_{C}+H_{R}+H_{B}$, where $H_{A}=h\sigma_{A}^{z}$, $H_{L}=k\sigma_{A}^{x}\sigma_{C_{1}}^{x}$, $H_{C}=k\sigma_{C_{1}}^{y}\sigma_{C_{2}}^{y}$, $H_{R}=k\sigma_{C_{2}}^{x}\sigma_{B}^{x}$, and $H_{B}=h\sigma_{B}^{z}$. Alice's projection and Bob's feedback unitary operators for the QET are defined respectively by
\begin{align}
P_{A}(n)&=\frac{1}{2}\left(I_{A}+n\vec{r}\cdot\vec{\sigma}_{A}\right), \\
U_{B}(n)&=I_{B}\cos\theta+in\vec{s}\cdot\vec{\sigma}_{B}\sin\theta ,
\end{align}
where the measurement result $n$ takes $n=\pm 1$, and $\vec{r}$ and $\vec{s}$ are unit vectors. These vectors and the angle $\theta$ are optimized so that the largest amount of work is extracted.

As a resource of QET, we prepare the lowest-energy state in the even parity sector of our Hilbert space (the eigenvalue of the parity operator $P=\sigma_{A}^{z}\sigma_{C_{1}}^{z}\sigma_{C_{2}}^{z}\sigma_{B}^{z}$ is $+1$). The state with the ground-state energy $\epsilon$ is then given by
\begin{align}
\left|\psi\right>=&Z\left[ \left|\downarrow\downarrow\uparrow\uparrow\right>+\left|\downarrow\uparrow\downarrow\uparrow\right>+\alpha\left(\left|\uparrow\uparrow\uparrow\uparrow\right>+\left|\uparrow\downarrow\downarrow\uparrow\right>\right) \right. \nonumber \\
& \left. +\left|\uparrow\uparrow\downarrow\downarrow\right>+\left|\uparrow\downarrow\uparrow\downarrow\right>+\beta\left(\left|\downarrow\downarrow\downarrow\downarrow\right>+\left|\downarrow\uparrow\uparrow\downarrow\right>\right)  \right] ,
\end{align}
where $\left|\uparrow\uparrow\uparrow\uparrow\right>=\left|\uparrow\right>_{A}\otimes\left|\uparrow\right>_{C_{1}}\otimes\left|\uparrow\right>_{C_{2}}\otimes\left|\uparrow\right>_{B}$, $\alpha=2k/(\epsilon+k-2h)$, $\beta=2k/(\epsilon+k+2h)$, $\epsilon=k\left(\alpha+\beta+1\right)$, and $Z^{2}\left(4+2\alpha^{2}+2\beta^{2}\right)=1$.

Important energetic quantities, $\Delta E_{B,B}$ and $\Delta E_{B,R}$, are evaluated as
\begin{align}
\Delta E_{B,B}=&\epsilon_{B}\left(1-s_{z}^{2}\right)\left(1-\cos 2\theta\right) \nonumber \\
&-h\left(r_{x}s_{y}C_{AB}-r_{y}s_{x}D_{AB}\right)\sin 2\theta , \\
\Delta E_{B,R}=&\epsilon_{R}\left(1-s_{x}^{2}\right)\left(1-\cos 2\theta\right)+kr_{x}s_{y}C_{AR}\sin 2\theta,
\end{align}
where various expectation values are defined as
\begin{align}
\epsilon_{B}&=\left<\psi\right|H_{B}\left|\psi\right>=2hZ^{2}\left(\alpha^{2}-\beta^{2}\right)\le 0, \\
\epsilon_{R}&=\left<\psi\right|H_{R}\left|\psi\right>=4kZ^{2}\left(\alpha+\beta\right)<0 , \\
C_{AB}&=\left<\psi\right|\sigma_{A}^{x}\sigma_{B}^{x}\left|\psi\right>=4Z^{2}\left(1+\alpha\beta\right)>0 , \\
D_{AB}&=\left<\psi\right|\sigma_{A}^{y}\sigma_{B}^{y}\left|\psi\right>=4Z^{2}\left(1-\alpha\beta\right)<0 , \\
C_{AR}&=\left<\psi\right|\sigma_{A}^{x}\sigma_{R}^{x}\sigma_{B}^{z}\left|\psi\right>=4Z^{2}\left(\alpha-\beta\right)>0 ,
\end{align}
and these two-point spin correlators are mutually connected through the following formula:
\begin{align}
kC_{AR}=h\left(C_{AB}-D_{AB}\right). \label{mutual}
\end{align}
The total extractable energy is given by $\Delta E_{B}=\Delta E_{B,B}+\Delta E_{B,R}$.

In the previous work, the parameters $\vec{r}$, $\vec{s}$, and $\theta$ that maximize $\Delta E_{B}$ have already been found. The parameters are taken as
\begin{align}
&\vec{r}^{\ast}=\left(\begin{matrix}0\cr 1\cr 0\end{matrix}\right) , \vec{s}^{\ast}=\left(\begin{matrix}1\cr 0\cr 0\end{matrix}\right) , \\
&\cos 2\theta^{\ast}=\frac{-\epsilon_{B}}{\sqrt{\epsilon_{B}^{2}+\left(hD_{AB}\right)^{2}}}, \\
&\sin 2\theta^{\ast}=\frac{hD_{AB}}{\sqrt{\epsilon_{B}^{2}+\left(hD_{AB}\right)^{2}}} ,
\end{align}
and $\Delta E_{B}^{\rm max}$ ($=\Delta E_{B}^{\ast}$) is given by
\begin{align}
\Delta E_{B}^{\rm max}=\epsilon_{B}+\sqrt{\epsilon_{B}^{2}+\left(hD_{AB}\right)^{2}},
\end{align}
where $\epsilon_{B}\le 0$. The formula of $\Delta E_{B}$ contains the factor $H_{i}-U_{B}^{\dagger}(n)H_{i}U_{B}(n)=U_{B}^{\dagger}(n)\left[U_{B}(n),H_{i}\right]$ ($i=B,R$), and thus we find that the interaction term $H_{R}=k\sigma_{R}^{x}\sigma_{B}^{x}$ commutes with $U_{B}(n)$ for the condition. As a result, we obtain $\Delta E_{B,R}^{\ast}=0$. On the other hand, we take $\vec{r}=(1,0,0)$ and $\vec{s}=(0,1,0)$ to maximize $\Delta E_{B,B}$. The maximization condition for $\Delta E_{B,B}$ is different from that for $\Delta E_{B}$.

\subsection{Positive $Q$ case (Maximization of $Q$)}

The maximum of $Q$ is given by
\begin{align}
Q^{\rm max}=&(-Q)^{\rm min}=\Delta E_{B,B}^{\rm max}-\Delta E_{B,B}^{\ast} \nonumber \\
=&\sqrt{\epsilon_{B}^{2}+\left(hC_{AB}\right)^{2}}-\sqrt{\epsilon_{B}^{2}+\left(hD_{AB}\right)^{2}}.
\end{align}
Since $C_{AB}>\left|D_{AB}\right|$, $Q^{\rm max}$ is positive. The nonlocal correlators $C_{AB}$ and $D_{AB}$ play a crucial role in the case of $Q^{\rm max}>0$. This result shows that the correlation behaves as heat. Note that the prefactor is $h$, not $k$, although $C_{AB}$ and $D_{AB}$ are finite in the presence of the many-body interaction.

\subsection{Maximization of $-Q$ (Minimization of $\Delta E_{B,B}$)}

The maximization condition of $-Q$ is equivalent to the minimization condition of $\Delta E_{B,B}$. Let us derive the condition. For this purpose, we first transform $\Delta E_{B,B}$ into the following form:
\begin{align}
\Delta E_{B,B}=&W\left(1-\cos 2\theta\right)+X\sin 2\theta \nonumber \\
=&W+\sqrt{W^{2}+X^{2}}\cos\left(2\theta+\delta\right) ,
\end{align}
where $W$ and $X$ are defined respectively by
\begin{align}
W&=\epsilon_{B}\left(1-s_{z}^{2}\right), \\
X&=-h\left(r_{x}s_{y}C_{AB}-r_{y}s_{x}D_{AB}\right),
\end{align}
and the phase factor $\delta$ is determined from the following relations:
\begin{align}
\cos\delta&=\frac{-W}{\sqrt{W^{2}+X^{2}}} , \\
\sin\delta&=\frac{-X}{\sqrt{W^{2}+X^{2}}}.
\end{align}
The minimum of $\Delta E_{B,B}$, $\Delta E_{B,B}^{\rm min}=\Delta E_{B,B}^{\times}$, is given as the lower bound of the following inequality
\begin{align}
W-\sqrt{W^{2}+X^{2}}\le\Delta E_{B,B}, \label{min}
\end{align}
where we take $2\theta+\delta=\pi$. The LHS of Eq.~(\ref{min}) is a function of $\vec{r}=(\sin\mu\cos\nu,\sin\mu\sin\nu,\cos\mu)$ and $\vec{s}=(\sin\xi\cos\eta,\sin\xi\sin\eta,\cos\xi)=(\cos\eta,\sin\eta,0)$. Clearly, $s_{z}=0$ for maximizing $\left|W\right|$ ($W<0$). This means $\xi=\pi/2$. The factor $X$ is represented as $X=-4Z^{2}h\sin\mu\sin\xi\left\{\sin(\eta-\nu)+\alpha\beta\sin(\eta+\nu)\right\}$, and we find that the optimized parameters are $\mu=\eta=\pi/2$ and $\nu=0$ (or $\pi$) for $\alpha\beta=(\epsilon-k)/(\epsilon+k)>0$.

Finally, $(-Q)^{\rm max}$ is given by
\begin{align}
\left(-Q\right)^{\rm max}=&\Delta E_{B,B}^{\ast}-\Delta E_{B,B}^{\times} \nonumber \\
=&\sqrt{\epsilon_{B}^{2}+\left(hD_{AB}\right)^{2}}+\sqrt{\epsilon_{B}^{2}+\left(hC_{AB}\right)^{2}},
\label{mQmax}
\end{align}
where the optimized parameters are defined by
\begin{align}
&\vec{r}^{\times}=\left(\begin{matrix}1\cr 0\cr 0\end{matrix}\right), \vec{s}^{\times}=\left(\begin{matrix}0\cr 1\cr 0\end{matrix}\right), \\
&\cos 2\theta^{\times}=\frac{\epsilon_{B}}{\sqrt{\epsilon_{B}^{2}+\left(hC_{AB}\right)^{2}}} , \\
&\sin 2\theta^{\times}=\frac{hC_{AB}}{\sqrt{\epsilon_{B}^{2}+\left(hC_{AB}\right)^{2}}}.
\end{align}
Note that $\vec{r}^{\times}$ and $\vec{s}^{\times}$ are consistent with those for the maximization condition of $\Delta E_{B,B}$ except for the difference of angle $\theta^{\times}$. This result indicates that the maximization protocol of $\Delta E_{B,B}$ generates a large amount of unavoidable heat. We can easily check that Eq.~(\ref{mQmax}) is positive. Furthermore, $(-Q)^{\rm max}$ is much larger than $\Delta E_{B}^{\rm max}=\sqrt{\epsilon_{B}^{2}+\left(hD_{AB}\right)^{2}}-\left|\epsilon_{B}\right|$. This feature has already been detected in our previous paper~\cite{Matsueda}, where even though $\Delta E_{B,B}^{\rm max}$ is achieved, $\Delta E_{B,R}$ is a large negative value, leading to the failure of energy extraction.

\subsection{Consistency between $(-Q)^{\rm max}$ and $\Pi^{\rm min}\left(\beta_{\rm eff}\right)$}

Here, we check the consistency between $(-Q)^{\rm max}$ and $\Pi^{\rm min}\left(\beta_{\rm eff}\right)$ in our Kitaev-like model. According to Eq.~(\ref{Pimin}), we need to evaluate $\bar{\rho}_{B}^{{\rm f}\times}$ and $\bar{\rho}_{B}^{{\rm f}\ast}$. Each component is evaluated as
\begin{align}
\bar{\rho}_{B}^{{\rm f}\times}=&\sum_{n}p_{n}^{\times}U_{B}^{\times}(n)\rho_{B}^{{\rm m}\times}(n)U_{B}^{\times\dagger}(n) \nonumber \\
=&\frac{1}{2}I_{B}+\left(\frac{\epsilon_{B}}{2h}\cos\left(2\theta^{\times}\right)+\frac{C_{AB}}{2}\sin\left(2\theta^{\times}\right)\right)\sigma_{B}^{z} , \\
\bar{\rho}_{B}^{{\rm f}\ast}=&\sum_{n}p_{n}^{\ast}U_{B}^{\ast}(n)\rho_{B}^{{\rm m}\ast}(n)U_{B}^{\ast\dagger}(n) \nonumber \\
=&\frac{1}{2}I_{B}+\left(\frac{\epsilon_{B}}{2h}\cos\left(2\theta^{\ast}\right)-\frac{D_{AB}}{2}\sin\left(2\theta^{\ast}\right)\right)\sigma_{B}^{z} ,
\end{align}
Finally, $\Pi^{\rm min}\left(\beta_{\rm eff}\right)=\Pi^{\times}$ is evaluated as
\begin{align}
\Pi^{\rm min}\left(\beta_{\rm eff}\right)=&\Pi^{\times} \nonumber \\
=&{\rm tr}_{B}\left(\bar{\rho}_{B}^{{\rm f}\times}H_{B}\right)-{\rm tr}_{B}\left(\bar{\rho}_{B}^{{\rm f}\ast}H_{B}\right) \nonumber \\
=&\epsilon_{B}\cos\left(2\theta^{\times}\right)+hC_{AB}\sin\left(2\theta^{\times}\right) \nonumber \\
&-\epsilon_{B}\cos\left(2\theta^{\ast}\right)+hD_{AB}\sin\left(2\theta^{\ast}\right) \nonumber \\
=&\sqrt{\epsilon_{B}^{2}+\left(hC_{AB}\right)^{2}}+\sqrt{\epsilon_{B}^{2}+\left(hD_{AB}\right)^{2}},
\end{align}
and this is exactly equal to $(-Q)^{\rm max}$ in Eq.~(\ref{mQmax}). Thus, the tightness of the upper bound of the Clausius inequality in Eq.~(\ref{Clausius2}) is correct for $\sigma_{B}\left(\beta_{\rm eff}\right)=\bar{\rho}_{B}^{{\rm m}\times}$.

\subsection{Effective Inverse Temperature}

We solve $\bar{\rho}_{B}^{{\rm m}\times}(n)=\sigma_{B}\left(\beta_{\rm eff}\right)$ to determine $\beta_{\rm eff}$. The LHS of this equation is given by
\begin{align}
\bar{\rho}_{B}^{{\rm m}\times}=&\sum_{n}p_{n}^{\times}\rho_{B}^{{\rm m}\times}(n)=\frac{1}{2}\left(I_{B}+\frac{\epsilon_{B}}{h}\sigma_{B}^{z}\right) \nonumber \\
=&2Z^{2}\left(\begin{matrix}1+\alpha^{2}&0\cr 0&1+\beta^{2}\end{matrix}\right).
\end{align}
We find that this is a diagonal matrix, and two eigenvalues are positive. To determine $\beta_{\rm eff}$, we compare the ratio of two eigenvalues in $\bar{\rho}_{B}^{{\rm m}\times}$ with that in $\sigma_{B}\left(\beta_{\rm eff}\right)$:
\begin{align}
e^{2\beta_{\rm eff}h}=\frac{h-\epsilon_{B}}{h+\epsilon_{B}}=\frac{1+\beta^{2}}{1+\alpha^{2}}\ge 1. \label{detbeta}
\end{align}
Then, we find $\beta_{\rm eff}\ge 0$, and in this case the effective temperature has a real and non-negative value. In the limit $h\rightarrow 0$, $\alpha=\beta$ and $\beta_{\rm eff}=0$. When the QET protocol cannot extract energy in the case of $h=0$, the effective temperature becomes infinity. Since Eq.~(\ref{detbeta}) gives physically reasonable inverse temperature, the final form of $\beta_{\rm eff}$ is given by
\begin{align}
\beta_{\rm eff}=\frac{1}{2h}\log\left(\frac{h-\epsilon_{B}}{h+\epsilon_{B}}\right).
\end{align}

Here, we discuss a possible alternative bound of $-Q$ on the basis of the positivity of $D\left(\bar{\rho}_{B}^{\rm f}||\sigma_{B}\right)$ in Eq.~(\ref{DD}). In this case, we require $\sigma_{B}\left(\beta_{\rm eff}^{\prime}\right)=\bar{\rho}_{B}^{{\rm f}\times}$ for the tight bound. This means that we regard the final state as a reference. The density matrix is represented as
\begin{align}
\bar{\rho}_{B}^{{\rm f}\times}=\left(\begin{matrix}\frac{h+\sqrt{\epsilon_{B}^{2}+\left(hC_{AB}\right)^{2}}}{2h}&0\cr 0&\frac{h-\sqrt{\epsilon_{B}^{2}+\left(hC_{AB}\right)^{2}}}{2h}\end{matrix}\right).
\end{align}
We clearly find $\left(\bar{\rho}_{B}^{{\rm f}\times}\right)_{\uparrow\uparrow}>0$. The component $\left(\bar{\rho}_{B}^{{\rm f}\times}\right)_{\downarrow\downarrow}$ is evaluated as
\begin{align}
\left(\bar{\rho}_{B}^{{\rm f}\times}\right)_{\downarrow\downarrow}=\frac{1}{2}\left(1-\sqrt{\frac{4+8\alpha\beta+\left(\alpha^{2}+\beta^{2}\right)^{2}}{4+8\left(\alpha^{2}+\beta^{2}\right)+\left(\alpha^{2}+\beta^{2}\right)^{2}}}\right),
\end{align}
and this is also positive. Unfortunately, because of $\left(\bar{\rho}_{B}^{{\rm f}\times}\right)_{\downarrow\downarrow}<\left(\bar{\rho}_{B}^{{\rm f}\times}\right)_{\uparrow\uparrow}$, we must introduce an inverted distribution or a negative inverse temperature. However, this is in some sense natural, since we now consider the process of heat generation inside of Bob's subsystem and the energy increases in the final state. In this case, the negative temperature is defined from
\begin{align}
e^{2\beta_{\rm eff}^{\prime}h}=&\frac{h-\epsilon_{B}\cos\left(2\theta^{\times}\right)-hC_{AB}\sin\left(2\theta^{\times}\right)}{h+\epsilon_{B}\cos\left(2\theta^{\times}\right)+hC_{AB}\sin\left(2\theta^{\times}\right)} \nonumber \\
=&\frac{h-\sqrt{\epsilon_{B}^{2}+\left(hC_{AB}\right)^{2}}}{h+\sqrt{\epsilon_{B}^{2}+\left(hC_{AB}\right)^{2}}}.
\end{align}

Based on the aforementioned careful point about the negativity of $\beta_{\rm eff}^{\prime}$, we derive a possible alternative bound. According to Eq.~(\ref{Clausius1}), we obtain
\begin{align}
&\beta_{\rm eff}^{\prime}Q+\Delta S_{B}-D\left(\bar{\rho}_{B}^{\rm m}||\sigma_{B}\right) \nonumber \\
&+\left[D\left(\bar{\rho}_{B}^{{\rm m}\ast}||\sigma_{B}\right)-D\left(\bar{\rho}_{B}^{{\rm f}\ast}||\sigma_{B}\right)\right]=-D\left(\bar{\rho}_{B}^{\rm f}||\sigma_{B}\right)\le 0,
\end{align}
and then we find
\begin{align}
-Q\le\Pi^{\prime},
\end{align}
where $\Pi^{\prime}$ is defined by
\begin{align}
\Pi^{\prime}=&\frac{1}{\beta_{\rm eff}^{\prime}}\left[\Delta S_{B}-D\left(\bar{\rho}_{B}^{\rm m}||\sigma_{B}\right)\right. \nonumber \\
&\;\;\;\;\;\;\;\;\left.+D\left(\bar{\rho}_{B}^{{\rm m}\ast}||\sigma_{B}\right)-D\left(\bar{\rho}_{B}^{{\rm f}\ast}||\sigma_{B}\right)\right] \nonumber \\
=&{\rm tr}_{B}\left(\bar{\rho}_{B}^{{\rm f}\times}H_{B}\right)-{\rm tr}_{B}\left(\bar{\rho}_{B}^{{\rm f}\ast}H_{B}\right).
\end{align}
This is essentially equal to $\Pi^{\rm min}\left(\beta_{\rm eff}\right)$ in Eq.~(\ref{Pimin}) and thus $(-Q)^{\rm max}$ in Eq.~(\ref{mQmax}). Therefore, although the definition of the reference state is different with each other, both of the positivity conditions for $D\left(\bar{\rho}_{B}^{\rm m}||\sigma_{B}\right)$ and $D\left(\bar{\rho}_{B}^{\rm f}||\sigma_{B}\right)$ finally give the same upper bound of $-Q$.

\section{Future Perspectives and Summary}
\label{summary}

We remark two future perspectives: (i) According to the definition of heat obtained in this study, the interaction $H_{R}$ connecting between Bob's subsystem and the rest of the whole system does not appear explicitly. How this relates to the open quantum system approach using the Lindblad equation, where the dissipator term is usually treated as heat, is an interesting topic for future research. (ii) It is still an open question whether Alice's projective measurement generates heat. In this paper, we also did not mention the first law of our effective thermodynamics due to the lack of the information about conserved quantities. These studies further proceed with thermodynamical perspectives behind the QET protocol.

In this paper, we constructed effective thermodynamics for the QET protocol by properly defining heat and work. The main point is to deeply consider the meaning of the passive state in the definition of the daemonic ergotropy: if the state is not passive even after the feedback unitary operation, there is still uncontrollable energy left in the system. This uncontrollable energy is attributed as heat. The heat is related to nonlocal correlation between Alice and Bob, and Bob can not directly access to this information. Thus, our definition of heat is consistent with the traditional meaning of heat as an uncontrollable energy. To complete the thermodynamical description of QET, we derived the generalized Clausius inequality including the effect of the quantum-state change due to the feedback unitary operation as well as the entropy change. We find that the direction of the inequality is reversed in comparison with the standard formula of the entropy production. We also introduced an explicit example of the QET model to examine the nature of heat. We proved the tightness of our Clausius inequality in the case that the maximum amount of heat remains in the system. The definitions of work and heat that we introduced here for the first time are expected to make a great contribution to the advances in modern quantum thermodynamics.

The author thanks Yusuke Masaki, Joji Nasu, Masahiro Takahashi, Miguel Lorenzo Laborte Ildesa, and Shota Arakaki for discussion. The author also thanks Hinata Yokoyama for giving several comments on the manuscript. This work was supported by the JSPS KAKENHI Grant Numbers JP24K06878, JP24K00563, JP24K02948, and CSIS in Tohoku University.


\begin{thebibliography}{99}
\bibitem{Strasberg}
P. Strasberg and A. Winter, Phys. Rev. X QUANTUM {\rm 2}, 030202 (2021).
\bibitem{Potts}
P. P. Potts, arXiv:2406.19206.

\bibitem{Ahmadi}
B. Ahmadi, S. Salimi, and A. S. Khorashad, Sicentific Reports {\bf 13}, 160 (2023).

\bibitem{Esposito}
M. Esposito, K. Lindenberg, and C. Van den Broeck, New J. Phys. {\bf 12}, 013013 (2010).

\bibitem{Allahverdyan}
A. E. Allahverdyan and T. M. Neiuwenhuizen, Phys. Rev. Lett. {\bf 85}, 1799 (2000).
\bibitem{Hilt}
S. Hilt and E. Lutz, Phys. Rev. A {\bf 79}, 010101(R) (2009).
\bibitem{Carrega}
M. Carrega, P. Solinas, M. Sassetti, and U. Weiss, Phys. Rev. Lett. {\bf 116}, 240403 (2016).
\bibitem{Alipour}
S. Alipour, F. Benatti, F. Bakhshinezhad, M. Afsary, S. Marcantoni, and A. T. Rezakhani, Sci. Rep. {\bf 6}, 35568 (2016).
\bibitem{Xu}
Y. Y. Xu, Phys. Rev. E {\bf 94}, 062145 (2016).
\bibitem{Bera}
M. Bera, A. Riera, M. Lewenstein, and A. Winter, Nat. Commun. {\bf 8}, 2180 (2017).
\bibitem{Llobet}
M. Perarnau-Llobet, H. Wilming, A. Riera, R. Gallego, and J. Eisert, Phys. Rev. Lett. {\bf 120}, 120602 (2018).
\bibitem{Strasberg2}
P. Strasberg, Phys. Rev. Lett. {\bf 123}, 180604 (2019).
\bibitem{Micadei}
K. Micadei, J. P. S. Peterson, A. M. Souza, R. S. Sarthour, I. S. Oliveira, G. T. Landi, T. B. Batalhao, R. M. Serra, and E. Lutz, Nature Comm. {\bf 10}, 2456 (2019).
\bibitem{Sapienza}
F. Sapienza, F. Cerisola, and A. J. Roncaglia, Nature Comm. {\bf 10}, 2492 (2019).
\bibitem{Dolatkhah}
H. Dolatkhah, S. Salimi, A. S. Khorashard, and S. Heseli, Sci. Rep. {\bf 10}, 9757 (2020).
\bibitem{Clesser}
J. D. Clesser and J. Anders, Phys. Rev. Lett. {\bf 127}, 250601 (2021).
\bibitem{Vallejo}
A. Vallejo, A. Romanelli, and R. Donangelo, Phys. Rev. E {\bf 103}, 042105 (2021).
\bibitem{Huang}
W.-M. Huang and W.-M. Zhang, Phys. Rev. A {\bf 106}, 032607 (2022).
\bibitem{Holdsworth}
T. Woldsworth and R. Kawai, Phys. Rev. A {\bf 106}, 062604 (2022).
\bibitem{Elouard}
C. Elouard and C. L. Latune, Phys. Rev. X QUANTUM {\bf 4}, 020309 (2023).
\bibitem{Dann}
R. Dann and R. Kosloff, New. J. Phys. {\bf 25}, 043019 (2023).
\bibitem{Bartosik}
P. Lipka-Bartosik, G. F. Diotallevi, and P. Bakhshinezhad, Phys. Rev. Lett. {\bf 132}, 140402 (2024).
\bibitem{Ye}
Y.-C. Ye, H.-G. Duan, and X.-T. Liang, Physica A {\bf 646}, 129869 (2024).


\bibitem{Aguilar}
M. Aguilar and E. Lutz, Sci. Adv. {\bf 11}, eadw8462 (2025).

\bibitem{Rivas}
\'{A}. Rivas, Phys. Rev. Lett. {\bf 124}, 160601 (2020).
\bibitem{Colla}
A. Colla and H.-P. Breuer, Phys. Rev. A {\bf 105}, 052216 (2022).
\bibitem{Seegebrecht}
A. Seegebrecht and T. Schilling, J. Stat. Phys. {\bf 191}, 34 (2024).
\bibitem{Colla2}
A. Colla and H.-P. Breuer, Quantum Sci. Technol. {\bf 10}, 015047 (2025).

\bibitem{Rupush}
W. Rupush and O. Gr\r{a}n\"{a}s, arXiv:2403.02022.
\bibitem{Zhou}
T. Zhou, J. Pu, and X. Wu, arXiv:2501.00832.

\bibitem{Allahverdyan2}
A. E. Allahverdyan, R. Balian, Th. M. Nieuwenhuizen, Europhys. Lett. {\bf 67}, 565 (2004).
\bibitem{Pusz}
W. Pusz and S. L. Woronowicz, Commun. math. Phys. {\bf 58}, 273 (1978).
\bibitem{Skrzypczyk}
P. Skrzypczyk, A. J. Short, and S. Popescu, Nat. Commun. {\bf 5}, 4185 (2014).
\bibitem{Llobet2}
M. Perarnau-Llobet, K. V. Hovhannisyan, M. Huber, P. Skrzypczyk, N. Brunner, and A. Ac\'{i}n, Phys. Rev. X {\bf 5}, 041011 (2015).
\bibitem{Francica}
G. Francica, J. Goold, F. Plastina, M. Paternostro, npj Quant. Inf. {\bf 3}, 12 (2018).
\bibitem{Bernards}
F. Bernards, M. Kleinmann, O. G\"{u}hne, and M. Paternostro, Entropy {\bf 21}, 771 (2019).
\bibitem{Francica2}
G. Francica, F. C. Binder, G. Guarnieri, M. T. Mitchison, J. Goold, and F. Plastina, Phys. Rev. Lett. {\bf 125}, 180603 (2020).
\bibitem{Lobejko}
M. Lobejko, Nat. Commun, {\bf 12}, 918 (2021).
\bibitem{Salvia}
R. Salvia and V. Giovannetti, Phys. Rev. A {\bf 105}, 012414 (2022).
\bibitem{Touil}
A. Touil, B. Cakmak, and S. Deffner, J. Phys. A: Math. Theor. {\bf 55}, 025301 (2022).
\bibitem{Hadipour}
M. Hadipour and S. Haseli, Sci. Rep. {\bf 14}, 24876 (2024).
\bibitem{Basu}
R. Basu, A. Chakrabortyy, H. Badhani, M. Alimuddin, and S. Bhattacharya, Phys. Rev. A {\bf 111}, 032416 (2025).

\bibitem{Hotta}
M. Hotta, Phys. Lett. A {\bf 372}, 5671 (2008).
\bibitem{Hotta2}
M. Hotta, Phys. Rev. D {\bf 78}, 045006 (2008).
\bibitem{Hotta3}
M. Hotta, J. Phys. Soc. Jpn. {\bf 78}, 034001 (2009).
\bibitem{Hotta4}
M. Hotta, Phys. Lett. A {\bf 374}, 3416 (2010).
\bibitem{Frey}
M. Frey, K. Funo, and M. Hotta, Phys. Rev. E {\bf 90}, 012127 (2014).
\bibitem{Trevison}
J. Trevison and M. Hotta, J. Phys. A: Math. Theor. {\bf 48}, 175302 (2015).
\bibitem{Rodriguez}
N. A. Rodriguez-Briones, H. Katiyar, E. Martin-Martinez, and R. Laflamme., Phys. Rev. Lett. {\bf 130}, 110801 (2023).
\bibitem{Ikeda}
K. Ikeda, Phys. Rev. Appl. {\bf 20}, 024051 (2023).
\bibitem{Wang}
J. Wang and S. Yao, Quantum {\bf 8}, 1564 (2024).
\bibitem{Hotta5}
M. Hotta and K. Ikeda, Quant. Inf. Proc. {\bf 24}, 186 (2025).
\bibitem{Matsueda}
H. Matsueda, Y. Masaki, K. Itoh, A. Ono, and J. Nasu, Phys. Rev. Res. {\bf 7}, 033137 (2025).
\bibitem{Itoh}
K. Itoh, Y. Masaki, and H. Matsueda, Phys. Rev. E {\bf 113}, 024108 (2026).

\bibitem{Sagawa}
T. Sagawa and M. Ueda, Phys. Rev. Lett. {\bf 100}, 080403 (2008).
\bibitem{Tajima}
H. Tajima, Phys. Rev. E {\bf 88}, 042143 (2013).
\bibitem{Park}
J. J. Park, K.-H. Kim, T. Sagawa, and S. W. Kim, Phys. Rev. Lett. {\bf 111}, 230402 (2013).
\bibitem{Funo}
K. Funo, Y. Watanabe, and M. Ueda, Phys. Rev. A {\bf 88}, 052319 (2013).
\bibitem{Manzano}
G. Manzano, F. Plastina, and R. Zambrini, Phys. Rev. Lett. {\bf 121}, 120602 (2018).
\bibitem{Minagawa}
S. Minagawa, M. H. Mohammadt, K. Sakai, K. Kato, and F. Buscemi, npj Quant. Inf. {\bf 11}, 18 (2025).

\bibitem{Abe1}
S. Abe and S. Okuyama, Phys. Rev. E {\bf 83}, 021121 (2011).
\bibitem{Abe2}
S. Abe, Phys. Rev. E {\bf 83}, 041117 (2011).
\bibitem{Abe3}
S. Abe and S. Okuyama, Phys. Rev. E {\bf 85}, 011104 (2012).
\bibitem{Bender}
C. M. Bender, D. C. Brody, and B. K. Meister, J. Phys. A: Math. Gen. {\bf 33}, 4427-4436 (2000).

\bibitem{Elouard2}
C. Elouard, D. A. Herrera-Mart\'{i}, M. Clusel, and A. Auff\`{e}ves, npj Quant. Inf. {\bf 3}, 9 (2017).
\bibitem{Yamamoto}
T. Yamamoto and Y. Tokura, Phys. Rev. Res. {\bf 6}, 013300 (2024).

\bibitem{Cafaro}
C. Cafaro and S. Mancini, Physica A {\bf 391}, 1610 (2012).
\bibitem{Vu}
Tan Van Vu and Keiji Saito, Phys. Rev. X {\bf 13}, 011013 (2023).
\bibitem{Nakazato}
M. Nakazato and S. Ito, Phys. Rev. Res. {\bf 3}, 043093 (2021).
\bibitem{Dechant}
A. Dechant and S. Sasa, J. Stat. Mech.: Theor. Exp., 063202, (2021).

\bibitem{Kitaev}
A. Kitaev, Ann. Phys. (NY) {\bf 321}, 2 (2006).
\bibitem{Nussinov}
Z. Nussinov and J. van den Brink, Rev. Mod. Phys. {\bf 87}, 1 (2015).
\bibitem{Hermanns}
M. Hermanns, I. Kimchi, and J. Knolle, Annu. Rev. Condens. Matter Phys. {\bf 9}, 17 (2018).


\end{thebibliography}
\end{document}